\documentclass[review]{elsarticle}

\usepackage{lineno,hyperref}
\usepackage{array}
\usepackage{appendix}
\usepackage{hyperref}
\usepackage{url}
\hypersetup{pdfauthor={Name}}
\usepackage[english]{babel}
\usepackage{color}
\usepackage{colorweb}
\usepackage{booktabs,caption}
\usepackage[flushleft]{threeparttable}
\usepackage{threeparttablex}
\usepackage{longtable}
\usepackage{dcolumn}
\usepackage{multirow}
\usepackage[htt]{hyphenat}
\usepackage{ulem}

\journal{Remote Sensing Applications: Society and Environment}

\biboptions{authoryear}

\begin{document}

\begin{frontmatter}

\title{Characterisation of night-time outdoor lighting in urban centres using cluster analysis of remotely sensed light emissions}

\author[1]{Máximo Bustamante-Calabria\corref{mycorrespondingauthor}}\ead{maximo@iaa.es}
\author[1]{Susana Martín-Ruiz}
\author[1,2]{Alejandro Sánchez de Miguel}
\author[1]{J. L. Ortiz}
\author[1]{J. M. Vílchez}
\author[1,3]{Jesús Aceituno}
\address[1]{Instituto de Astrofísica de Andalucía (IAA-CSIC). Granada, Spain}
\address[2]{Dpto. de Física de la Tierra y Astrofísica. Universidad Complutense, Madrid, Spain}
\address[3]{Observatorio Astronómico de Calar Alto (CAHA). Almería, Spain}

\begin{abstract}
Evidence of the negative impact of light pollution on ecosystems is increasing every year. Its monitoring and study requires the identification, characterisation and control of the emitting sources. This is the case of urban centres with outdoor lighting that spills light outside the place it is intended to illuminate. The quantity and nature of the pollutant (artificial light at night) depends on the lamps used and how they are positioned, which is important because a greater proportion of blue light means a greater scattering effect. In this study, we analysed the emissions of 100 urban centres in the north of Granada province (Spain), using images from International Space Station (ISS) in 2012 and 2021, and compared the results with public lighting inventories and with data from the VIIRS instrument. Using inference and cluster analysis techniques, we confirmed an overall increase in emissions and a shift in their colour towards blue, consistent with the results of the lighting inventory analysis. However, there is a discrepancy with the results obtained from VIIRS images, which is explained by the lack of sensitivity of this instrument to blue light. We concluded that the analysis of ISS images is a powerful tool for the study and characterisation of street lighting and its evolution, especially when the changes occur in the blue band.  
\end{abstract}

\begin{keyword}
outdoor lighting \sep light pollution \sep remote sensing \sep cluster analysis 
\end{keyword}

\end{frontmatter}

\section{Introduction}

One approach to quantifying light pollution and monitor its evolution is to measure the background brightness of the night sky at the zenith. This is usually done from the ground using photometric equipment \citep{Cinzano2005, SanchezdeMiguel2017, BustamanteCalabria2021}. The other method of measuring artificial light at night is to use remote sensing from orbiting sensors. DMSP (Defense Meteorological Satellite Program) was the first instrument used to obtain nighttime images, which produced the first global nighttime map in 1989. This instrument, along with the Operational Limscan System (OLS), allowed NOAA to produce calibrated 1 km resolution global maps between 1996 and 2010 \citep{Elvidge1999}. For example, the analysis of DMSP-OLS data has made possible to estimate the evolution of energy consumption in public lighting \citep{SanchezdeMiguel2014}. Since 2011, the Visible Infrared Imaging Radiometer Suite (VIIRS) instrument on the Suomi NPP satellite, which has a panchromatic sensor that also measures nighttime light, has been in operation \citep{Hillger2013, Liao2013}. It provides calibrated data with a spatial resolution of 740 metres without saturation in urban areas, making it one of the most widely used data sources for the study of light pollution (some examples in \cite{EstradaGarcia2016, Levin2017, Roman2018} and the recent review of \cite{Combs2023}). The main drawback of this instrument lies in the fact that it does not detect the blue emission peak of LED lamps, so as LEDs replace sodium vapour in public outdoor lighting, VIIRS underestimates the brightness of urban areas \citep{Levin2020}. There are also several commercial Chinese satellites that take high-resolution multispectral night-time images \citep{Zheng2018, Guo2023}. 

Satellite images are often calibrated in radiance units and are useful as a basis for models. The best known is Falchi's light pollution map based on VIIRS data \citep{Falchi2016}. On the other hand, as \cite{SanchezdeMiguel2020} show, the diffuse brightness present in these images corresponds to the sky brightness at zenith measured on the ground, so these remote data can be used as a map of sky brightness at zenith. In addition to this instrument, another source of orbital night-time imagery is astronauts' images from the International Space Station (ISS), which are more comprehensive across the visible spectrum because they are obtained with conventional photographic cameras \citep{SanchezdeMiguel2019}. Although this source of data lacks the regularity offered by satellites, the image base is large enough to have several images for an area in different years. This has enormous potential for studying the growing problem of light pollution \citep{SanchezdeMiguel2014a}.

Since both public and private outdoor lighting at night is the main source of light pollution, it is essential to characterise and monitor it. This is possible if up-to-date public lighting inventories are available, but these usually exclude ornamental lighting and private infrastructure lighting. Thus, the analysis of remote colour images can be a very useful tool in these studies.

This research aims to asnwer the following questions:

\begin{itemize}
	\item How reliable is the characterisation of light pollution sources (outdoor lighting at night) using data extracted from images taken from the ISS?
	\item Have light emissions to the sky (type and quantity) changed significantly in recent years?
	\item Has the generalized trend of switching to LED lamps reduced or increased light pollution? And in terms of colour, how have light emissions changed?
	\item Can the populations centers be characterized or grouped according to their light emissions measured from space? Are the results consistent with public lighting inventories?
	\item And finally, to what degree private lighting is contributing to the problem?
\end{itemize}

\section{Study area}

\begin{figure}[ht]
\resizebox{\hsize}{!}{\includegraphics{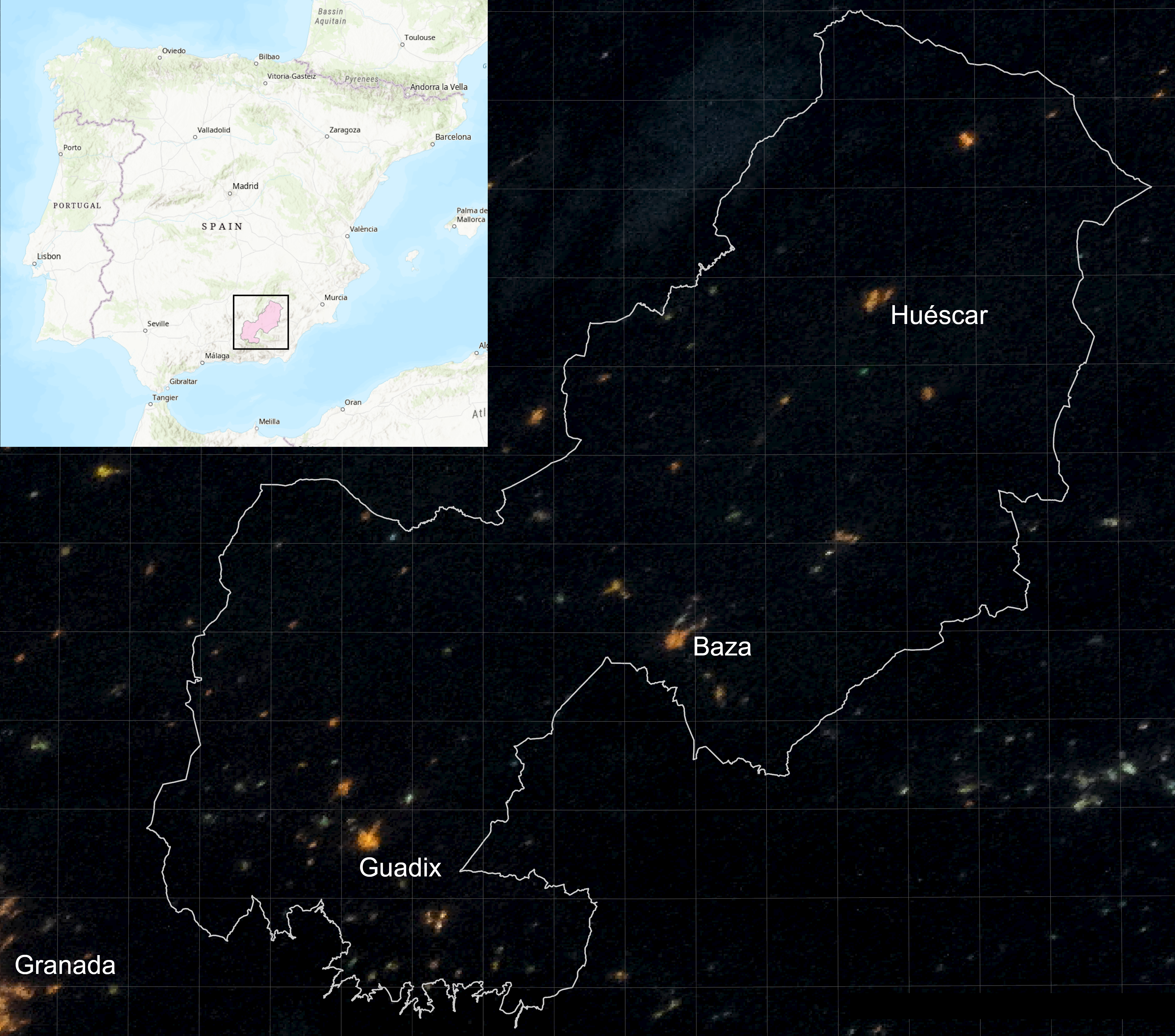}}
\caption{Study area: Granada Geopark (map base: ESRI, NASA)}
\label{fig:01}
\end{figure}

The area covered by this work corresponds to the so-called Granada Geo\hyp{park}, located in the north of the province of Granada (figure \ref{fig:01}). Thanks to its landscapes and geological singularities, the Granada Geopark is part of the European Geoparks Network and was declared a UNESCO Global Geopark in July 2020. The conservation of the natural night sky is one of the objectives of the administrations responsible for its management, and to achieve this, it is essential to study and monitor light emissions from urban centres. 

This territory includes 47 municipalities in the shires of Huéscar, Baza, Guadix, El Marquesado and Los Montes. Among these, there is one munici\hyp{pality} with more than 20000 inhabitants (Baza) and another close by (Guadix), followed by Huéscar with around 7000 inhabitants. In the range between 2000 and 5000 inhabitants are Cúllar, Caniles, Benalúa, Benamaurel, Castril, Puebla de Don Fadrique, Purullena, Valle del Zalabí and Zújar. The remaining 35 municipalities have fewer than 2000 inhabitants, and 10 have fewer than 500 \citep{INE2022}. The urban centres of this area are compact. Landscaped housing estates are rare and streets are generally narrow and treeless, except for a few main roads.

This area is among the darkest places remaining in mainland Spain according to sky brightness maps\footnote{\url{https://www.lightpollutionmap.info}} \footnote{\url{https://pmisson.users.earthengine.app/view/trends}}, while at the same time having very good overall weather and high mountains well suited for professional astronomy activities.

\section{Data and methods}

The methodology of this work has been divided into three parts: The first phase consisted in the pre-processing of the images taken by the astronauts of the International Space Station, the second is to obtain the values of the emissions by colour band in the different population nuclei of the area under study, and the third in the statistical analysis of the emission data. For the first two phases we used the ArcGIS Pro software in its advanced version ArcInfo \citep{ESRI2022}. For the third phase we worked with the R language in the RStudio environment \citep{RStudio2022}.

\subsection{Remote data}

The source data are raw images captured by astronauts from the International Space Station \citep{NASA2022}. Photos taken on 6 February 2012 (at 4:50 GMT) and 24 July 2021 (at 23:35 GMT) were used for this study, covering the northern part of the province of Granada (see Table \ref{tab:ISSdata} for the photo data). These images have been calibrated, corrected and georeferenced according to the methodology of \cite{SanchezdeMiguel2021}, resulting in images that provide radiance values for each pixel and band in nanowatts per angstrom, steradian and square centimetre ($nW \cdot \textup{\r{A}}^{-1}  sr^{-1}  cm^{-2}$). The process was as follows:

\begin{enumerate}
	\item Decoding: the raw data is converted into standard format (tiff) with the RGB channels isolated. 
	\item Linearity corrections: corrects the non-linear behaviour of the camera at high photon counts.  
	\item Flat field/vignetting correction: the uneven illumination of the sensor due to the camera lens is corrected. 
	\item Spectral channel calibration: compensates for variations in spectral sensitivity between the camera's channels. 
	\item Astrometric calibration: reference stars are identified on calibration ima\hyp{ges} and their measurements are used for absolute calibration of the central images. 
	\item Georeferencing: ground control points within the central images are identified and used to transform them into georeferenced coordinates. 
	\item Photometric calibration: within the calibration images, the known absolute flux of stars is determined to find the instrumental constant for converting photon counts in the central images to International System units. 
	\item Radiometric adjustment: the instrumental constant derived from the stars is applied to the central images and adjusted according to the camera settings used during imaging. 
	\item Atmospheric correction: other remaining elements, including aerosols and Raleigh scattering, are adjusted.
\end{enumerate}

The next step was to delineate each urban nucleus with polygons. In this way, only the urban areas are included in the calculations.

Finally, since the surface area of the pixel is known, the conversion from ($nW \cdot \textup{\r{A}}^{-1}  sr^{-1}  cm^{-2}$) to ($W \cdot \textup{\r{A}}^{-1}  sr^{-1}$) units has been done. In this way, is obtained the sum of the radiances per band for the polygon that delimits each of the agglomerations.

The analysis of light emissions has been done per population centre or infrastructure detected in the image, adding up to a total of 100 units (87 population centres and 13 private infrastructures).

\begin{table}[]
\begin{center}
\caption{ISS image data}
\scriptsize
\label{tab:ISSdata}
\begin{tabular}{@{}lccccc@{}}
\toprule
\textit{Image ID} & \textit{Date} & \textit{GMT} & \textit{Camera/Focal Length} & \textit{Alt (km)} & \textit{Tilt angle} \\ 
\midrule
iss030e086586 & 2012-02-06 & 04:50:15 & Nikon D3S/85 mm & 393 & 17 \\
iss030e086587 & 2012-02-06 & 04:50:16 & Nikon D3S/85 mm & 393 & 16 \\
iss030e086588 & 2012-02-06 & 04:50:20 & Nikon D3S/85 mm & 393 & 7 \\
iss030e086589 & 2012-02-06 & 04:50:24 & Nikon D3S/85 mm & 393 & 23 \\
iss030e086590 & 2012-02-06 & 04:50:25 & Nikon D3S/85 mm & 393 & 23 \\
iss030e086591 & 2012-02-06 & 04:50:30 & Nikon D3S/85 mm & 393 & 18 \\
iss065e200633 & 2021-07-24 & 23:34:55 & Nikon D5/24 mm & 415 & 14 \\
\bottomrule
\end{tabular}
\end{center}
\end{table}

On the other hand, we have analysed data from the Day/Night Band of the Visible Infrared Imaging Radiometer Suite instrument (VIIRS DNB) to make a comparison with the results obtained from the ISS images. We have calculated radiance data per municipality from 2012 to 2021 using data from the Earth Observation Group\footnote{Data available on the websites \url{https://eogdata.mines.edu/products/vnl/} and \\ \url{https://lighttrends.lightpollutionmap.info}}, based on the papers of \cite{Elvidge2017VIIRS} and \cite{Coesfeld2020a}.

\begin{figure}[ht]
\begin{center}
\includegraphics[scale=0.4]{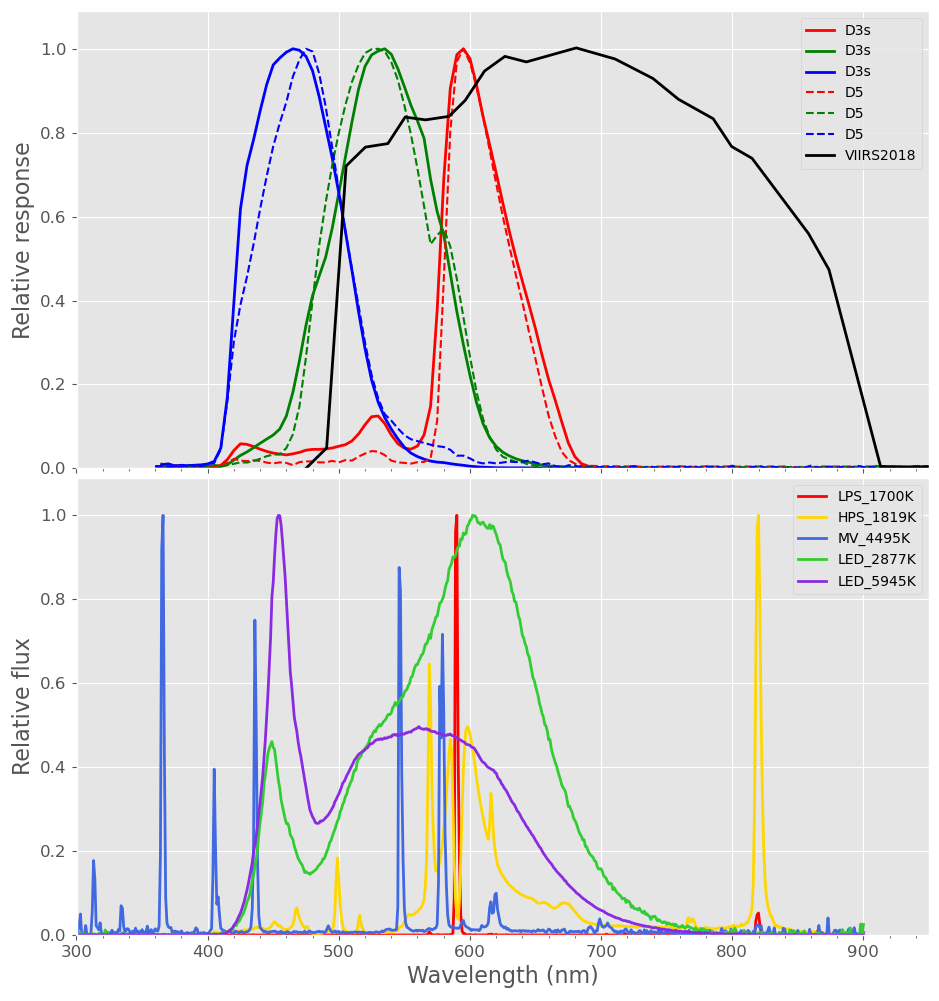}
\caption{Up: Spectral responses of the Nikon D3s (colours solid line) and Nikon D5 (colours dashed line) cameras and VIIRS satellite sensor (black solid line). Bottom: Spectral fluxes of the most common streetlighting types of different temperatures in Kelvin.}
\label{fig:02}
\end{center}
\end{figure}

Figure \ref{fig:02} shows the spectral responses of the sensors on the cameras that took the images from the ISS and the VIIRS-DNB instrument \citep{SanchezdeMiguel2019}, compared with the emission spectra of the most common lamps used in street lighting \citep{Tapia2017}. The emission of high-pressure sodium lamps (HPS) is perfectly captured by VIIRS (including the infrared peak), whereas this sensor is completely blind to the blue emission peak of LED lamps. However, DSLR camera sensors have a spectral response that better covers the emission of LED lamps.

\subsection{Public lighting inventories}

The inventory provided by Diputacion de Granada\footnote{Geographic information system of the province of Granada, available at \url{https://ide.dipgra.es/siggra_ide/index}} has led to establish a correspondence between the results of the analysis of colour indices from remote images and those of public lighting. In the resulting database were found 47 records (one per municipality) and 27 variables (providing information on power, light points, lamp types, upper hemispheric flux, etc.). In addition, three new variables have been created to help characterise this lighting, namely the percentage of warm lamps, the percentage of cold lamps and the percentage of degraded lamps. High and low pressure sodium lamps have been considered as warm lamps; LEDs (with a colour temperature above 2200 K)\footnote{This classification of warm and cool lamps is not related to the manufacturer's classification, but takes into account the presence of some emission peaks at wavelengths shorter than 500 nm.}, mercury vapour and metal halide lamps have been considered as cool lamps; and mercury vapour, metal halide and fluorescent lamps have been considered as degraded lamps. Most of the database provided was updated in 2020 and 2021.

\subsection{Statistical analysis}

As we are only interested in the light emission from nuclei and infrastructure, it was necessary to identify and demarcate them using polygons. Subsequently, geostatistical tools were used to calculate the sum of the value of the pixels included in the polygon of each nucleus and for each colour channel. This value is the total radiance of the population nucleus in $W \cdot \textup{\r{A}}^{-1}  sr^{-1}$. 

In this way, a database for each detected nucleus and infrastructure was created, containing as variables the total emissions for each year (2012 and 2021) in red, green and blue, the colour indices B/G (the ratio between blue and green emissions) and G/R (the ratio between green and red emissions) and the population. This database (associated to the vectorial layer of geographical information of the polygons of nuclei and infrastructures) is the starting point for the statistical analysis in R. The process has been as follows:

\begin{enumerate}
	\item Data preparation and preliminary analysis. Calculation of signal-to-noise ratio (SNR) for each population centre, discarding cases with SNR $< 3$. 
	\item Development of linear regression models for the colour indices and each year. Detection of outliers and investigation of atypical and high leverage records. 
	\item Inference analysis between emission values and colour indices for each year to searching for statistically significant differences. 
	\item K-means cluster analysis based on the colour indices obtained from ISS images for 2021 and for the difference between 2021 and 2012. This ana\hyp{lysis} identifies groups of populations with similar characteristics in the current colour of their emissions, as well as groups of nuclei with similar behaviour in the spectral evolution of their emissions from 2012 to 2021. For the cluster analysis, the variables were first scaled with respect to the mean. After checking for clustering tendencies and calculating the distance matrices between the points, a total of 26 statistical indices were calculated to evaluate six clustering assumptions. The number of clusters chosen was the one that performed best on most of the indices.
	\item K-means cluster analysis of the public lighting inventory of the municipa\hyp{lities} to evaluate whether groups can be obtained that correspond to the previous ones in point 4. This analysis has been carried out  taking into account only the types of lamps grouped according to the colour of the light they emit.
	\item K-means cluster analysis of the municipalities' differences in blue emissions obtained from the ISS images and the differences in emissions obtained from the VIIRS instrument data. 
	\item Finally, the groups obtained from the colour index analysis of the ISS data are compared with those obtained from the street lighting analysis and with those obtained from the VIIRS data to see if there is any correspondence between them.  
\end{enumerate}

As support for the writing of the R code, \cite{Crawley2012}, \cite{Wikle2019} and \cite{Carvalho2023} were consulted.

\section{Results}

After the preliminary analysis, the nuclei belonging to the municipalities of Huéneja, Dólar and Ferreira are discarded, because they are affected by high clouds in the image corresponding to 2021, as well as the nuclei with SNR $< 3$. A total of 79 nuclei remained for analysis (72 urban, 2 industrial and 5 private). The complete list can be found in \ref{A:1}, Table \ref{tab:emin_nucleos}.

\subsection{Comparative analysis of emissions and colour indices between 2012 and 2021}

In the analysis of the ISS images, the question is whether there are significant differences in both emissions and colour indices between 2012 and 2021. As these are measurements of the same cores at different times, it is necessary to consider related samples. The main problem found in assessing the assumptions is related to the non-normality of the data: only the values of the B/G index are close to a normal distribution, but neither the G/R index nor the emission values are. When analysing the differences between the two years, the same result is obtained. For this reason, a non-parametric Wilcoxon signed-rank test is applied.\footnote{The Wilcoxon signed-rank test compares the positions of the values of two related samples.}.

\begin{figure}[ht]
\resizebox{\hsize}{!}{\includegraphics{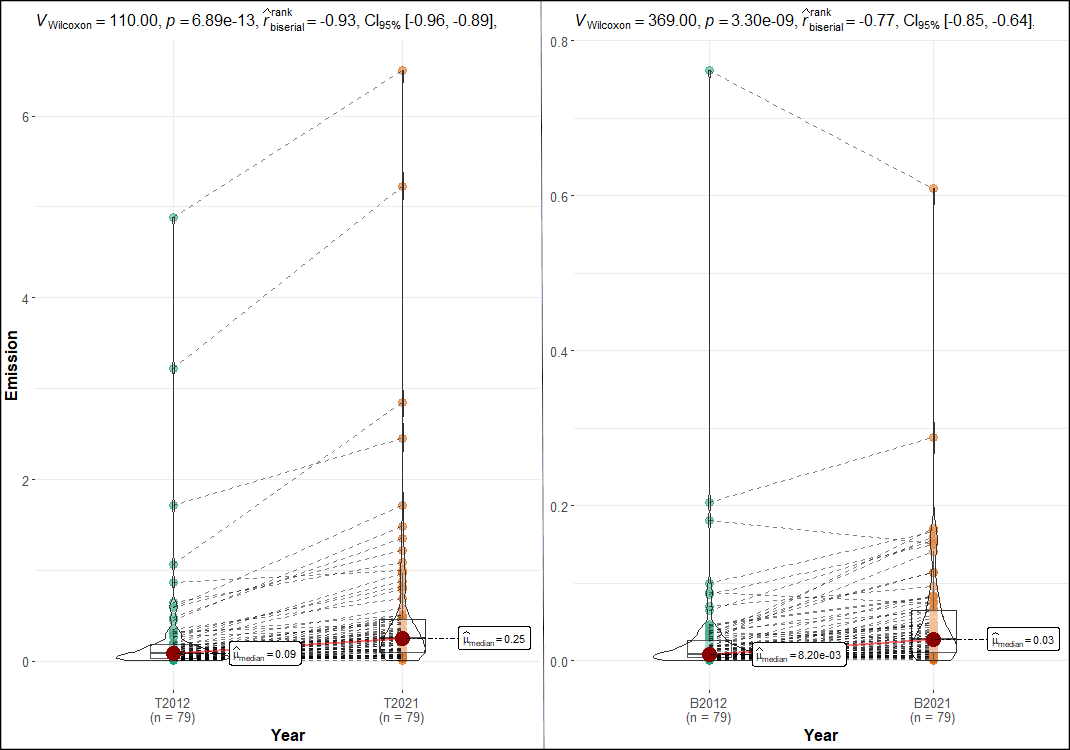}}
\caption{Result of Wilcoxon test for total emissions (left) and blue band emissions (right) from 79 population centres and infrastructures. Emission units in $W \cdot \textup{\r{A}}^{-1} sr^{-1}$.}
\label{fig:03}
\end{figure}

\begin{figure}[ht]
\resizebox{\hsize}{!}{\includegraphics{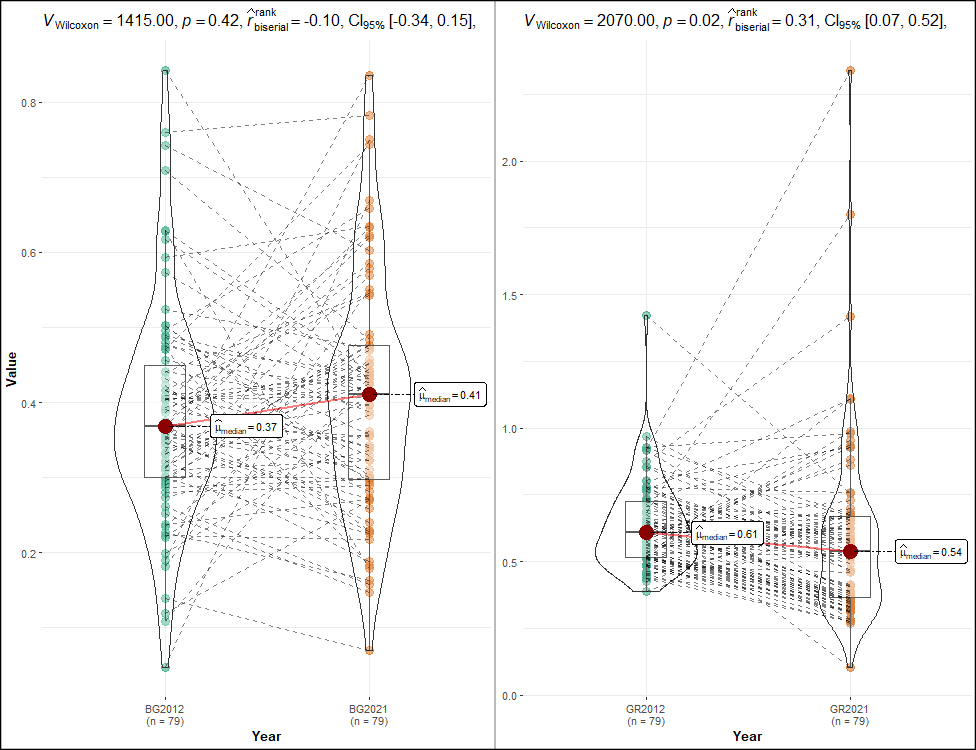}}
\caption{Result of Wilcoxon test for B/G index (left) and G/R index (right) of 79 population centres and infrastructures.}
\label{fig:04}
\end{figure}

The results can be summarised as follows: 

\begin{itemize}
	\item There is a statistically significant increase of 0.16 ($\pm$ 0.01) $W \cdot \textup{\r{A}}^{-1} sr^{-1}$ in measured emissions in all colour bands from 2012 to 2021 (figure \ref{fig:03}, left), which represents an overall increase of 178\% and an annual increase of 19.7\%. 
	\item There is a statistically significant increase of 0.022 ($\pm$ 0.001) $W \cdot \textup{\r{A}}^{-1} sr^{-1}$ in emissions in the blue band from 2012 to 2021 (figure \ref{fig:03}, right), which represents an overall increase of 266\% and an annual increase of 30\%.
	\item There is a increase (not statistically significant) in the median B/G colour index from 2012 to 2021 (figure \ref{fig:04}, left). The median increase in the B/G index is 0.04 ($\pm$ 0.05). No clear conclusion can be drawn because the absolute error is greater than the value of the difference. 
	\item There is a decrease (not statistically significant) in the median G/R colour index from 2012 to 2021 (figure \ref{fig:04}, right). The median decrease in the G/R index is 0.07 ($\pm$ 0.08). No clear conclusion can be drawn because the absolute error is greater than the value of the difference.
\end{itemize}

\subsection{Cluster analysis of population centres according to the colour indexes of their emissions}

An analysis of the comparative results between 2012 and 2021 for the colour indices shows that they are very uneven between the agglomerations. The differences in the medians are not statistically significant and the errors do not allow any clear conclusions to be drawn. Cluster analysis techniques can be used to see if there are groups of nuclei that behave similarly in terms of the colour of their light emissions. After verifying that there is a tendency for clustering, k-means partitioning is applied. The detailed results can be found in \ref{A:3}.

\subsubsection{Cluster analysis for 2021 colour indices}

\begin{figure}[ht]
\resizebox{\hsize}{!}{\includegraphics{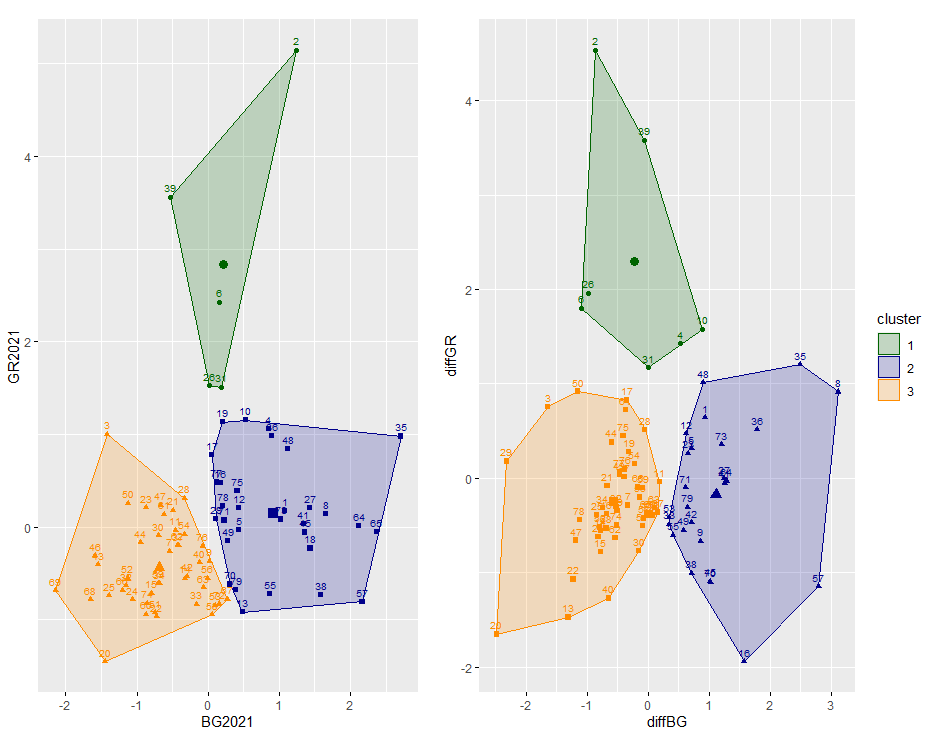}}
\caption{Partitioning by k-means according to the colour indices (BG2021 and GR2021) in the year 2021 (left) and according to the difference in the colour indices (diffBG and diffGR) between the years 2021 and 2012 (right). Values are adimensional and scaled to the mean value. The full list of each group can be found in table \ref{tab:kmeans}, \ref{A:3}.}
\label{fig:05}
\end{figure}

The best option is to divide into 3 groups using the k-means method (figure \ref{fig:05}, left). The B/G index turned out to be the most important variable for this grouping.

\begin{itemize}
	\item Group 1. It consists of 5 population centres. It is characterised by B/G index values centred around the average, but well above it in the G/R index. It is an odd group, characterised by significant green emissions. However, the B/G ratio is not below average (and in some cases even above average), so they must be emitting quite a lot of blue band light. If a direct relationship with the type of lighting is assumed, mercury vapour, fluorescent or metal halide lamps (which degrade over time and emit a greenish light) and white LEDs (responsible for blue) should be present. 
	\item Group 2. This group includes 27 agglomerations and 4 private infrastructures. It is characterised by an above-average B/G index, while the G/R index is in the average, so that blue emission predominates. These would be nuclei or installations that have replaced most of their lamps with LEDs. 
	\item Group 3. With 43 members, this is the largest group in which the rest of the nuclei or facilities studied are located. It consists of 42 urban nuclei and 1 private infrastructure. This group has the warmest emissions, with B/G and G/R indices lower than average. It is to be expected that sodium vapour lamps still predominate in their lighting.
\end{itemize}

\subsubsection{Cluster analysis for differences in colour indices between 2021 and 2012}

Based on the analysis of the similarities according to the colour of their light emissions for the most recent data of the cores studied, and considering that the global analysis does not produce a statistically significant result for the changes in the colour indices between 2012 and 2021, it is proposed to study whether there are changes by group and whether they behave differently. To do this, a cluster analysis is used, where the variables are the differences in the values of the colour indices between the two years.

Analogous to the 2021 case, the best option is to divide into 3 groups using the k-means method (figure \ref{fig:05}, right). The most important variable leading to this grouping is the difference in the B/G index.

\begin{itemize}
	\item Group 1 is composed of 7 agglomerations. It is characterised by a signifi\hyp{cant} increase in the G/R index, while the B/G remains more or less the same.
	\item Group 2 is made up of 24 nuclei and 1 private infrastructure. They are characterised by a significant increase in the B/G index from 2012 to 2021, together with a slight decrease in the G/R index, which corresponds to an increase in blue light emissions and a decrease in the proportion of green light. This probably means that they have already replaced the majority of their lamps with LEDs. 
	\item Group 3. 43 agglomerations and 4 private infrastructures belong to this group. Their colour indices have decreased very slightly compared to the average between 2012 and 2021. For public lighting, this could be explained by the initial replacement of mercury vapour and fluorescent lamps by sodium vapour, which is currently being replaced by LEDs, but these do not yet represent a significant proportion of the total. 
\end{itemize}

\subsubsection{Inference analysis for colour indices between 2012 and 2021 considering grouping}

\begin{figure}[]
\resizebox{\hsize}{!}{\includegraphics{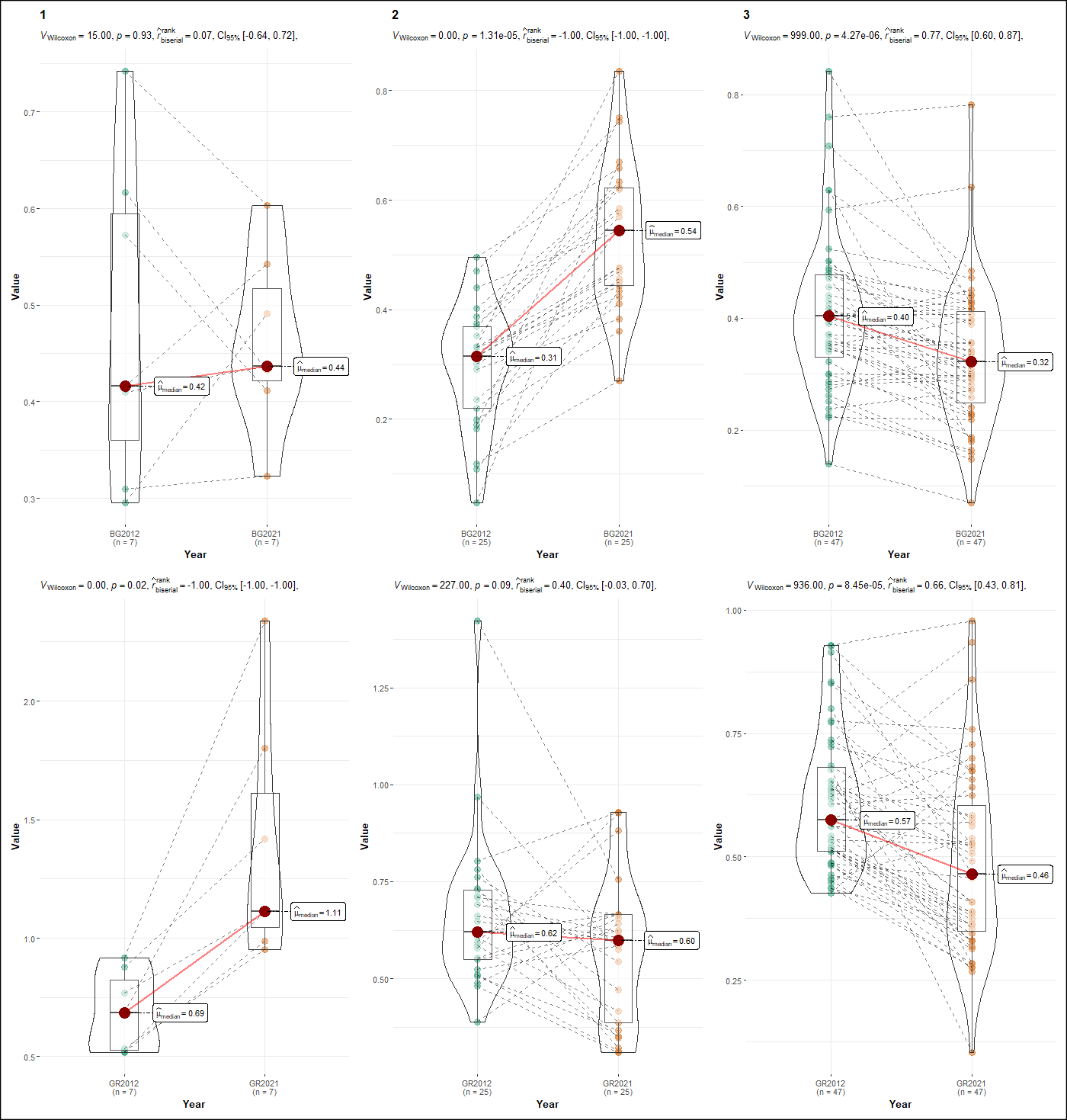}}
\caption{Result of Wilcoxon test for the three groups resulting from the k-means partitioning based on the differences in colour indices between the years 2021 and 2012. Columns: Group 1 (left), Group 2 (centre) and Group 3 (right). Rows: B/G index (top) and G/R index (bottom). Index values are dimensionless.}
\label{fig:06}
\end{figure}

A non-parametric Wilcoxon signed-rank test is applied to compare the colour indices between 2012 and 2021, taking into account the division into groups obtained in the previous subsection (Figure \ref{fig:06}).

\begin{itemize}
	\item Group 1: consisting of 7 nuclei with no statistically significant differences in the B/G index, but with significant differences in the G/R index. The median G/R index increased by 0.42 ($\pm$ 0.17).
	\item Group 2: consisting of 25 nuclei that do not show statistically significant differences in the G/R index, but do show differences in the B/G index. The median B/G index increased by 0.23 ($\pm$ 0.07).
	\item Group 3: consisting of 47 nuclei with statistically significant differences in the two indices. The median of the B/G index decreased by 0.08 ($\pm$ 0.04) and the median of the G/R index decreased by 0.11 ($\pm$ 0.06).
\end{itemize}

Group 2 (nuclei that have increased their blue emissions) shows the most significant differences and the smallest relative error.

\subsection{Cluster analysis based on lamp type}

The sum of the photons arriving directly from the emitting lamps, those reflected by the surfaces, those scattered by the atmosphere, etc. is the light emission captured by a remote sensor in a urban centre. It is therefore to be expected that there will be a relationship between the detected colour and the spectral characteristics of the emitters. In order to check the consistency between the results obtained from the analysis of the ISS images and the cha\hyp{racteristics} of the outdoor lighting, the public lighting inventory provided by the Diputación de Granada is analysed. 

A cluster analysis is therefore performed using the percentage of warm, cold and degraded lamps as variables. This has a more direct relationship with the colour of the emissions. In this case, the best option is to divide into 3 groups using the k-means method (figure \ref{fig:07}, left). 

\begin{itemize}
	\item Group 1: 11 municipalities with a predominance of degraded lamps (mercury vapour, fluorescent, halide). They maintain old lamps or have replaced discharge lamps with fluorescent or other types, but there are hardly any LEDs or sodium vapour lamps. 
	\item Group 2: 9 municipalities with a clear predominance of cold lamps, while warm and degraded lamps are below average. They have almost completed switching to white LEDs.
	\item Group 3: 20 municipalities with a predominance of sodium vapour lamps.
\end{itemize}

\subsection{Emissions with VIIRS data and cluster analysis based on blue emission differences measured from ISS images and VIIRS emission differences}

\begin{figure}[ht]
\resizebox{\hsize}{!}{\includegraphics{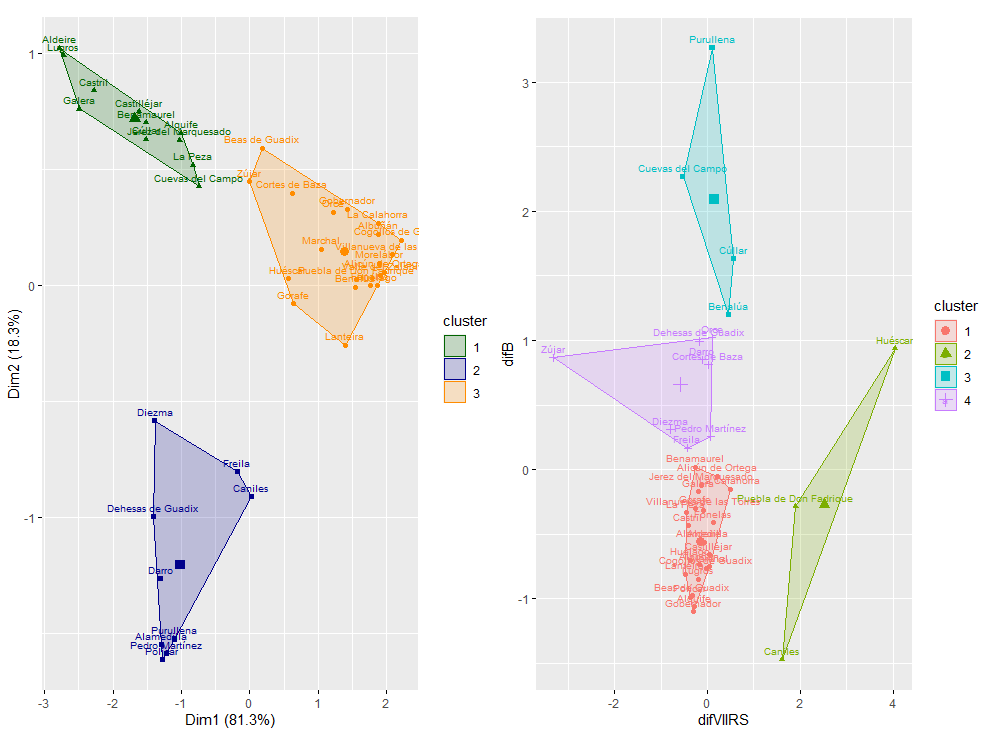}}
\caption{Left: K-means partitioning according to the characteristics of the lamps in the street lighting inventory. The three variables (proportion of warm, cold and degraded lamps in the inventory) were scaled to two dimensions. Right: K-means partitioning according to blue emission differences measured from ISS images (difB) and VIIRS emission differences (difVIIRS) between 2021 and 2012. The full list of each group can be found in table \ref{tab:kmeans}, \ref{A:3}.}
\label{fig:07}
\end{figure}

To compare the results with data from the VIIRS instrument, emissions per municipality are calculated for the years 2012 and 2021 (the mean of the available data set for the year). Table \ref{tab:EMmunicipios} in \ref{A:1} shows the behaviour per municipality compared to the emissions measured from the ISS images. In addition, a cluster analysis is performed taking into account the differences between 2021 and 2012 in the emissions per municipality according to VIIRS, together with the differences in the blue band according to ISS images. The aim of this analysis is to check the coherence between the data obtained by the two sensors. Figure \ref{fig:07} (right) shows the k-means partitioning applied to the municipalities, resulting in 4 groups. The most influential variable in this partition is the difference in blue emissions, while for most municipalities it results in a similar variation in VIIRS. There is no similarity with the groups obtained in the other cluster analyses. Detailed data can be found in table \ref{tab:kmeans} (\ref{A:3}).

However, there are three main situations based on the differences in emissions:

\begin{itemize}
	\item 11 municipalities where a decrease in emissions is observed according to VIIRS data, but with an increase in the blue band according to ISS ima\hyp{ges}. This can be explained by the replacement of sodium vapour lamps with cold light LEDs (with a blue emission peak invisible to the VIIRS instrument).
	\item 3 municipalities where a decrease in emissions is observed according to VIIRS and in the blue band according to ISS imagery. This case can be explained by the replacement of mercury vapour lamps, the maintenance of sodium vapour lamps and an improvement in luminaires.
	\item 30 municipalities with a small increase in emissions according to VIIRS, accompanied by a increase in blue according to ISS images. This is the most common situation. The explanation could be related to a switch to warm white LEDs with an increase in intensity.
\end{itemize}

\section{Discussion}

\subsection{Matching results}

There should be a correspondence between the groups obtained from the analysis of the remote images in terms of colour and those obtained according to the predominant type of lamp. Since the main town is the one with the highest concentration of public lighting points, its emissions are the most representative of the municipality, so a correlation between the two sets of data (colour indices from remote sensing and lamp type from inventories) can be made for at least the most important population centres. A predominance of warm lamps should result in a low B/G index, and vice versa: a predominance of cold lamps implies a high B/G index. The presence of degraded lamps would be indicated by a predominance of green, so having a low B/G index together with a high G/R.

\begin{figure}[ht]
\resizebox{\hsize}{!}{\includegraphics{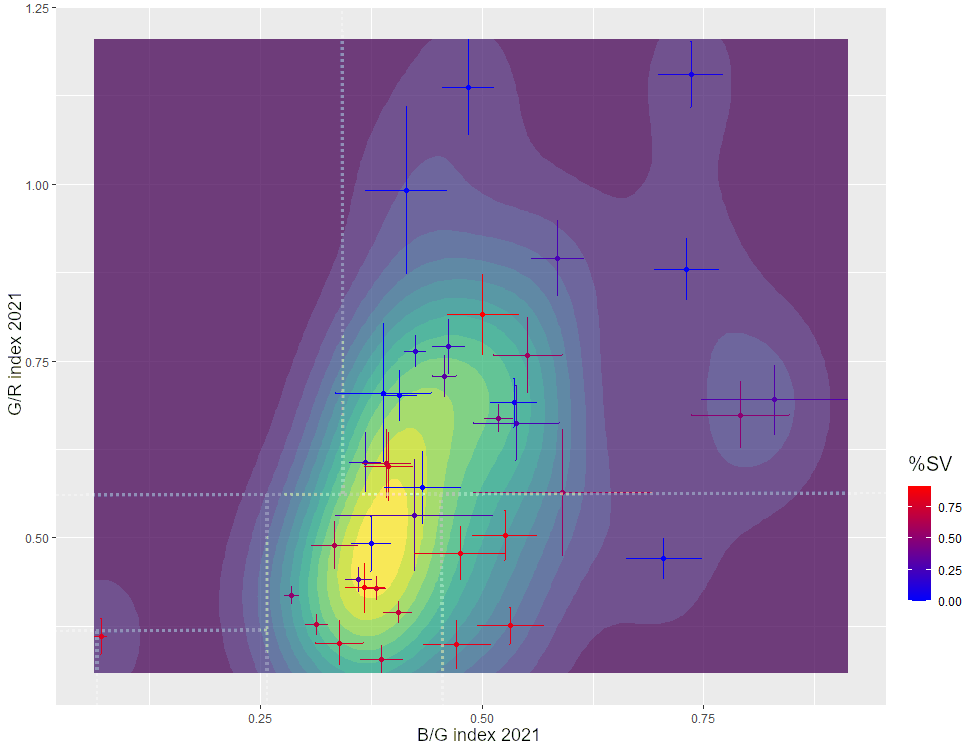}}
\caption{Correlation between B/G and G/R indices calculated from ISS images for the municipalities capitals of Granada Geopark (2021). The dot's colour represents the percentage of sodium lamps (\%SV) in the public lighting inventory. Background colour represents dot density (towards yellow, more density of dots). The dashed lines represent the sectors of \cite{SanchezdeMiguel2019} (the upper right sector corresponds to the cold light lamps). Indices are adimensional.}
\label{fig:08}
\end{figure}

Figure \ref{fig:08} shows the municipalities capitals according to the B/G and G/R colour indices of their emissions calculated from the 2021 ISS images. There is a moderate ($\rho = 0.44$) and statistically significant correlation between the two indices, as well as between the G/R index and the prevalence of sodium vapour lamps in public lighting ($\rho = -0.58$), represented by the colour of the dots. If this graph is compared with the one obtained by \cite{SanchezdeMiguel2019} for the indices of the different types of lamps used in public lighting, an analogous relationship can be seen, although in this case it is more scattered. This is normal, given that the ISS images do not have sufficient resolution to distinguish emissions from ornamental and private lighting in these urban centres, so it is not possible to subtract them from public lighting. In addition, while Sánchez de Miguel's graph shows direct measurements from laboratory lamps, the light captured by the camera from the ISS was scattered by the Earth's atmosphere. 

However, it must be taken into account that the inventory data may have changed in the year before the date of the last ISS image (2021). The change may always be in the same direction: replacement of sodium vapour or degraded lamps with LEDs. As of today, we are not aware of any municipalities installing amber LEDs, which could cause warmer emissions. On the contrary, cold white LEDs (4000 K in many cases), or more recently 3000 K or slightly warmer, have been installed everywhere. But in no case has there been a return to sodium vapour or a lamp that produces light with almost no blue emission, although this would be desirable. 

The municipalities with a predominance of cold lamps and few degraded lamps in their inventory belong to group 2 of those obtained by the k-means method (9 municipalities). There should be some correspondence with group 2 when looking at the groups obtained by colour index analysis in 2021, with higher B/G and moderate G/R indices. To see if there is a correspondence between the groups with predominantly green light and the municipalities with degraded lamps, it is necessary to look at group 1 of municipalities to compare them with group 1 for the 2021 colour indices. Only two cases share a group. 

However, it is important to note that we have assumed that all the lamps we have labelled as degraded are actually degraded, and this is not necessarily the case. Mercury vapour lamps are often old and give a greenish light, but other types may be newer and still give a cold light without a predominance of green. Thus, if the comparison is made between the groups of cold light and degraded lamps and the groups of blue-dominant and green-dominant lamps, a greater degree of similarity can be observed: 12 out of 20 (60\%) municipalities belonging to groups 1 and 2 (cold light or degraded lamps) correspond to their capitals in group 2 obtained by the k-means method according to the colour index differences (agglomerations that have shifted their emissions more towards blue).

For the municipalities with the highest share of sodium vapour lamps, there is a good correspondence with the municipal capitals that emit the least in the blue band between 2012 and 2021 (lowest B/G and G/R ratios). 17 out of 20 municipalities (85\%) belonging to group 3 (warm light lamps) correspond to their capitals in group 3 obtained by the k-means method according to the colour indices in 2021.

\subsection{Contribution of private infrastructure to light emissions}

In terms of private infrastructure, two large installations (the Andasol solar thermal power plant\footnote{\url{https://www.power-technology.com/projects/andasolsolarpower/}} and the logistics centre\footnote{\url{https://www.google.es/maps/place//@37.3365478,-3.0996392,15z}}) emit in the blue band more than a town of almost 20000 inhabitants such as Guadix. In the area studied, 5 private infrastructures and 2 industrial areas are responsible for 12\% of the total light emissions detected in 2021.

\subsection{Comparison of results with VIIRS data}

The results obtained with the ISS images do not agree with those obtained with VIIRS. Municipalities with a significant increase in blue emissions in the ISS images have not increased (or even decreased) their emissions according to the VIIRS data. In most cases there is a small increase of less than 5\% according to VIIRS, while according to ISS data there is an increase of 20 to 50\% (and there are even cases of more than 100\%) from the year 2021 compared to 2012. Looking at Figure \ref{fig:02} and the results of the cluster analysis according to colour indices and their correspondence with public lighting, this discrepancy can be explained.

Figure \ref{fig:02} shows that the VIIRS sensor is completely blind to the blue peak of LED lamps, while at the same time being sensitive to the infrared emission peak of high pressure sodium lamps. Thus, in municipalities that have replaced sodium vapour lamps with cold LEDs during this period, VIIRS will pick up an apparent decrease in emissions because the infrared peak of the HPS is no longer produced, while it will not pick up the new blue peak of the LED. This effect is more pronounced the colder the LED (and the higher the peak at 460nm). VIIRS data are therefore found to be unreliable for assessing the evolution of light emissions when switching from HPS lamps to cool LEDs.

\subsection{On the limitations of ISS data}

The limitations of ISS images are related to the lack of control over the time and conditions in which they are taken, since the astronauts do not follow a criterion or methodology designed for scientific use, but rather for recreational or educational purposes. They are mainly:

\begin{itemize}
	
	\item Scarcity and temporal irregularity of available imagery for an area. This makes it difficult to obtain statistically robust temporal trends, as is possible with satellite data.
	\item Insufficient spatial resolution, which means that it's difficult to distinguish the light emitted by private elements or installations located within the agglomerations, as well as by ornamental lighting. It must therefore be taken into account that the emissions calculated for each agglomeration are not only due to public lighting.
	\item Differences in the angle to the Earth's surface. While the satellites have a zenithal perspective, the astronauts take the images from any angle. This means that the light received by a city looks different from different perspectives.
\end{itemize}

Temporal irregularity means that attention must be paid to the existence of changing conditions that may affect urban emissions. For example, seasonal variations in vegetation (due to deciduous trees) or the presence of snow in winter. On the other hand, hourly variations need to take into account changes related to human activity, ornamental lighting or programmed variations in lighting intensity.

The images used for this work correspond to different seasons (February and July). However, snow falls occasionally in the study area and there is no snow in the February image. The presence of trees is also not a factor, as there are few green areas in the population centres studied. As for the time difference, one image was taken at 4:50 GMT (winter) and the other at 23:34 GMT (summer), which is 5:50 (GMT+1) and 1:34 (GMT+2) in official Spanish time. Thus, the two images are after 1:00, when some of the ornamental lighting is usually switched off (although some ornamental lighting remains on all night). 

\subsection{Comparison with other research results}

The increase in blue light emissions found in this paper is consistent with the results of other studies, both based on ground-based photometric measurements and remotely sensed image analysis. \cite{Robles2021} estimated a 32\% increase in Johnson B filter brightness for Madrid between 2015 and 2019. On the other hand, the study by \cite{SanchezdeMiguel2022} for several European cities (based on the analysis of ISS images) shows a clear increase in blue light emissions associated with the conversion of street lighting to LEDs. In our case, the median increase in blue emissions between 2012 and 2021 for the 79 cases studied was 266\% (figure \ref{fig:03}).

In terms of characterising street lighting through remote imaging, there is the work of \cite{Rybnikova2021}. They build statistical models from spectral data to predict the contribution of lamp types to the brightness and colour of each pixel in the image. The study focuses on Haifa (Israel) and achieves a good predictive capability for sodium vapour and metal halide lamps. In our case, the main difference is the small size and heterogeneity of the nuclei studied, so we found cluster analysis techniques more useful. We also obtained a good prediction for both cold and sodium vapour lamps.

Finally, regarding the contribution of domestic lighting to light pollution, \cite{Garcia_Gil_Gonzalez_Dorta_2016} used lighting models to estimate the contribution of domestic lighting to the luminous flux in an area of Barcelona; they concluded that it could be responsible for half of the light emissions emitted to the sky. Similar conclusions were reached by \cite{Bara2019} for A Coruña and Arteixo using photometric methods. In 2021, \cite{Kyba2021} concluded from a controlled experiment in Tucson (Arizona) that street lighting was only responsible for between 16\% and 21\% of the emissions in that city. In our study, we were not able to distinguish private sources within urban centres, but we were able to confirm the important contribution of private installations to the light pollution of an area. This is related to the way in which warehouses, industrial estates, etc. are lit all night long, with unjustified intensities and directions. 

\section{Conclusions}

In this research, we analysed data on street lighting inventory and emissions from urban centres from remotely sensed imagery, comparing data separated by 9 years. We conclude that it is possible to characterise night-time lighting by analysing ISS imagery. K-means cluster analysis techniques revealed an acceptable degree of similarity between groups with a high B/G index and a predominance of cold light lamps in their inventory (60\% agreement between groups). There is also a high degree of similarity between the groups with a low B/G index and a predominance of sodium vapour lamps (85\% agreement between the groups).

Regarding the evolution of light pollution from 2012 to 2021, the inference analysis points to a clear increase, especially in blue band, which is consistent with the evolution of the public lighting inventory, where there has been a general replacement of sodium vapour lamps by white LEDs in a group of nuclei that we identified quite clearly using cluster analysis of the colour index values.

In addition, two facts stand out: on the one hand, the shift towards blue emissions, although general, is more pronounced in small towns; on the other hand, private infrastructure contributes at least 12\% of total artificial light emissions, not counting the contribution of the private sector within urban areas.

We can therefore conclude that the analysis of ISS images is a powerful tool for the study and characterisation of street lighting and its evolution, especially since the VIIRS instrument is not sensitive to blue light, but it is subject to some drawbacks that make us exercise due caution in the interpretation of the results. Given the lack of control over the conditions under which the images are taken and the uncertain future of the ISS missions, there is a greater need than ever to put into orbit instruments that provide data from the night side of the Earth that are suitable for scientific use, sensitive to the blue band of the spectrum and accessible to the research community. 

\section*{Funding}

Authors acknowledge financial support from the grant CEX2021-001131-S funded by MCIN/AEI/ 10.13039/501100011033 and from project PID2020-112789GB-I00. This publication also is part of the project PDC2022-133985-I00, funded by MCIN/AEI/10.13039/501100011033 and by the European Union "NextGenerationEU"/PRTR. This research was also funded by Diputación Pro\hyp{vincial} de Granada, as part of a comprehensive study of night sky quality in Granada Geopark.

\section*{Data availability statement}

The data used for this research and detailed maps can be found at: \url{https://doi.org/10.5281/zenodo.7951019}

\section*{Description of author's responsibilities}

M.B.C., S.M.R. and A.S.M. conceived the study; A.S.M. conducted the cali\hyp{bration} procedures; S.M.R. made the management, permits and data request; S.M.R., J.L.O. and J.M.V. conducted the funding requests; M.B.C. performed the statistical and GIS analysis; M.B.C. wrote the original manuscript; S.M.R., J.L.O., J.M.V. and J.A. proofread the English translation. All authors reviewed the manuscript.

\section*{Conflicts of interest}

The authors declare no conflict of interest.

\section*{Declaration of Generative AI and AI-assisted technologies}

The authors did not use generative AI and AI-assisted technologies in the writing of this paper.

\bibliography{sample}

\appendix
\clearpage
\appendixpage

\section{Emissions and indexes per municipality and population centre}\label{A:1}

\begin{ThreePartTable}

\begin{TableNotes}
\scriptsize
\item Light emissions measured from ISS images, total (T) and in blue band (B), for the years 2012 and 2021 (numbers refer to the total for each municipality, units in $W \cdot sr^{-1} \textup{\r{A}}^{-1}$). $\Delta T$: annual increase in total emissions. $\Delta B$: annual increase in blue emissions. $\Delta VIIRS$: annual increase in total emissions according to VIIRS imagery.
\end{TableNotes}

\scriptsize
\begin{longtable}{@{}lccccccc@{}}
\caption{Total light emissions and in the blue band of the municipalities of Granada Geopark for the years 2012 and 2021.}
\scriptsize
\label{tab:EMmunicipios} \\
\toprule
MUNICIPALITY & T 2012 & T 2021 & B 2012 & B 2021 & $\Delta T$ (\%) & $\Delta B$ (\%) & $\Delta VIIRS$ (\%)\\ 
\midrule
\endfirsthead

\multicolumn{8}{c}{{\bfseries \tablename\ \thetable{}:}  continuation}\\
\addlinespace[0.9ex]
\toprule
MUNICIPALITY & T 2012 & T 2021 & B 2012 & B 2021 & $\Delta T$ (\%) & $\Delta B$ (\%) & \ $\Delta VIIRS$ (\%)\\
\midrule    
\endhead

\midrule
\endfoot
\bottomrule
\insertTableNotes
\endlastfoot 

Alamedilla & 0.0182 & 0.1898 & 0.0028 & 0.0199 & 104.72 & 68.13 & 3.17 \\
Albuñán & 0.0636 & 0.2774 & 0.0054 & 0.0166 & 37.32 & 23.42 & 1.61 \\
Aldeire & 0.1652 & 0.3135 & 0.0283 & 0.0455 & 9.97 & 6.77 & 2.31 \\
Alicún de Ortega & 0.0202 & 0.3647 & 0.0023 & 0.0367 & 189.6 & 166.09 & 12.04 \\
Alquife & 0.0956 & 0.1668 & 0.0111 & 0.0137 & 8.28 & 2.57 & 0.38 \\
Baza & 5.3529 & 6.9874 & 0.8260 & 0.6586 & 3.39 & -2.25 & -3.24 \\
Beas de Guadix  & 0.0401 & 0.0621 & 0.0028 & 0.0059 & 6.09 & 12.48 & -1.84 \\
Benalúa & 0.4521 & 1.4854 & 0.0377 & 0.1142 & 25.39 & 22.54 & 0.73 \\
Benamaurel & 0.3457 & 0.7009 & 0.0562 & 0.1192 & 11.42 & 12.46 & 0.99 \\
Calahorra, La & 0.1437 & 0.4590 & 0.0174 & 0.0484 & 24.37 & 19.79 & 1.34 \\
Caniles & 0.8222 & 1.4253 & 0.1084 & 0.1177 & 8.15 & 0.96 & 2.14 \\
Castilléjar & 0.1619 & 0.3456 & 0.0197 & 0.0335 & 12.6 & 7.81 & 1.12 \\
Castril & 0.1771 & 0.4889 & 0.0158 & 0.0647 & 19.57 & 34.34 & -1.04 \\
Cogollos de Guadix & 0.0566 & 0.1944 & 0.0064 & 0.0166 & 27.02 & 17.88 & 3.65 \\
Cortes de Baza & 0.1522 & 0.7675 & 0.0195 & 0.1184 & 44.92 & 56.43 & 1.99 \\
Cortes y Graena & 0.1039 & 0.1766 & 0.0124 & 0.0192 & 7.77 & 6.03 & 0.47 \\
Cuevas del Campo & 0.2804 & 0.9578 & 0.0298 & 0.1482 & 26.85 & 44.08 & -1.66 \\
Cúllar & 0.7902 & 1.7716 & 0.0917 & 0.2680 & 13.8 & 21.34 & 0.69 \\
Darro & 0.1764 & 0.5787 & 0.0193 & 0.0840 & 25.34 & 37.26 & 0.12 \\
Dehesas de Guadix & 0.0303 & 0.3411 & 0.0047 & 0.0742 & 113.91 & 163.2 & 3.05 \\
Diezma & 0.1879 & 0.2335 & 0.0204 & 0.0670 & 2.7 & 25.29 & -5.09 \\
Fonelas & 0.0684 & 0.4843 & 0.0046 & 0.0268 & 67.58 & 53.64 & 2.19 \\
Freila & 0.2768 & 0.3667 & 0.0350 & 0.0767 & 3.61 & 13.24 & 1.29 \\
Galera & 0.1257 & 0.2644 & 0.0190 & 0.0493 & 12.25 & 17.71 & -0.17 \\
Gobernador & 0.1097 & 0.2182 & 0.0050 & 0.0040 & 10.99 & -2.25 & -0.05 \\
Gor & 0.1406 & 0.0929 & 0.0368 & 0.0276 & -3.77 & -2.79 & -1.26 \\
Gorafe & 0.0252 & 0.1445 & 0.0061 & 0.0322 & 52.73 & 47.19 & -1.04 \\
Guadix & 4.1074 & 6.7110 & 0.3161 & 0.4482 & 7.04 & 4.64 & 0.21 \\
Huélago & 0.0682 & 0.2441 & 0.0076 & 0.0201 & 28.64 & 18.25 & -1.21 \\
Huéscar & 1.3842 & 3.7481 & 0.1382 & 0.2381 & 18.97 & 8.03 & 4.38 \\
Jerez del Marquesado & 0.0674 & 0.2921 & 0.0081 & 0.0401 & 37.04 & 44.1 & 1.79 \\
Lanteira & 0.1553 & 0.2641 & 0.0191 & 0.0279 & 7.78 & 5.09 & -1.82 \\
Lugros & 0.0080 & 0.0492 & 0.0009 & 0.0085 & 57.29 & 94.09 & 6.06 \\
Marchal & 0.0333 & 0.2668 & 0.0023 & 0.0130 & 77.86 & 51.86 & 1.46 \\
Morelábor & 0.1179 & 0.4880 & 0.0098 & 0.0286 & 34.89 & 21.47 & 0.55 \\
Orce & 0.4808 & 1.3567 & 0.0430 & 0.1135 & 20.24 & 18.22 & 0.86 \\
Pedro Martínez & 0.0466 & 0.4560 & 0.0046 & 0.0493 & 97.7 & 108.05 & 3.4 \\
Peza, La & 0.1203 & 0.2553 & 0.0140 & 0.0495 & 12.46 & 28.07 & -1.12 \\
Polícar & 0.0154 & 0.0523 & 0.0029 & 0.0057 & 26.54 & 10.81 & -1.38 \\
Puebla de Don Fadrique & 0.6816 & 2.0337 & 0.0851 & 0.1147 & 22.04 & 3.87 & 1.47 \\
Purullena & 0.2227 & 0.8129 & 0.0272 & 0.1728 & 29.44 & 59.52 & 1.16 \\
Valle del Zalabí & 0.1577 & 0.6853 & 0.0198 & 0.0800 & 37.18 & 33.77 & 1.41 \\
Villanueva de las Torres & 0.0259 & 0.2893 & 0.0028 & 0.0282 & 112.81 & 102.47 & 3.76 \\
Zújar & 0.8719 & 1.0048 & 0.0850 & 0.1502 & 1.69 & 8.53 & -5.05 \\ 
\bottomrule
\end{longtable}
\end{ThreePartTable}

\begin{ThreePartTable}

\begin{TableNotes}
\scriptsize
\item B/G and G/R colour indexes and errors calculated from ISS images of the year 2021. Indexes are dimensionless.
\end{TableNotes}

\scriptsize
\begin{longtable}{@{}l|l|cccc@{}}
\caption{B/G and G/R colour indexes and errors by centre or installation} \label{tab:emin_nucleos} \\
\toprule
\textbf{MUNICIPALITY} & \textbf{CENTRE} & \textbf{B/G 2021} & \textbf{e} & \textbf{G/R 2021} & \textbf{e} \\
\midrule
\endfirsthead

\multicolumn{6}{c}{{\bfseries \tablename\ \thetable{}:}  continuation}\\
\addlinespace[0.9ex]
\toprule
\textbf{MUNICIPALITY} & \textbf{CENTRE} & \textbf{B/G 2021} & \textbf{e} & \textbf{G/R 2021} & \textbf{e} \\
\midrule    
\endhead

\midrule
\endfoot
\bottomrule
\insertTableNotes
\endlastfoot

ALAMEDILLA & Alamedilla & 0.43 & 0.04 & 0.57 & 0.05 \\
\midrule
ALBUÑAN & Albuñán & 0.48 & 0.05 & 0.48 & 0.04 \\
\midrule
ALDEIRE & Aldeire & 0.70 & 0.04 & 0.47 & 0.03 \\
\midrule
ALICÚN DE ORTEGA & Alicún de Ortega & 0.53 & 0.04 & 0.38 & 0.03 \\
\midrule
ALQUIFE & Alquife & 0.42 & 0.09 & 0.53 & 0.08 \\
\midrule
\multirow{4}{6em}{BAZA} & Río de Baza & 0.48 & 0.12 & 0.28 & 0.07 \\
& Baza & 0.35 & 0.01 & 0.41 & 0.01 \\
& Baúl & 0.18 & 0.06 & 0.10 & 0.02 \\
& P.I. el Baico & 0.47 & 0.02 & 0.37 & 0.01 \\
\midrule
BEAS DE GUADIX & Beas de Guadix & 0.59 & 0.10 & 0.56 & 0.09 \\
\midrule
BENALUA & Benalúa & 0.38 & 0.01 & 0.43 & 0.02 \\
\midrule
\multirow{2}{6em}{BENAMAUREL} & Benamaurel & 0.46 & 0.02 & 0.77 & 0.04 \\
& San Marcos & 0.63 & 0.07 & 0.52 & 0.07 \\
\midrule
\multirow{3}{6em}{CANILES} & Caniles & 0.36 & 0.01 & 0.44 & 0.02 \\
& La Vega & 0.44 & 0.04 & 0.98 & 0.09 \\
& Los Gallardos & 0.19 & 0.02 & 0.34 & 0.03 \\
\midrule
\multirow{2}{6em}{CASTILLEJAR} & Castilléjar & 0.37 & 0.02 & 0.61 & 0.04 \\
& & & & & \\
\midrule
\multirow{3}{6em}{CASTRIL} & Castril & 0.37 & 0.02 & 0.49 & 0.04 \\
& Almontaras & 0.49 & 0.05 & 0.99 & 0.12 \\
& Fátima & 0.34 & 0.04 & 0.58 & 0.06 \\
\midrule
COGOLLOS DE GUADIX & Cogollos de Guadix & 0.50 & 0.04 & 0.82 & 0.06 \\
\midrule
\multirow{4}{6em}{CORTES DE BAZA} & La Teja & 0.43 & 0.17 & 1.41 & 0.24 \\
& Los Laneros & 0.67 & 0.05 & 0.64 & 0.05 \\
& Cortes de Baza & 0.55 & 0.04 & 0.76 & 0.05 \\
& Campo Cámara & 0.33 & 0.03 & 0.66 & 0.04 \\
\midrule
\multirow{4}{6em}{CORTES Y GRAENA} & Graena & 0.26 & 0.07 & 0.54 & 0.05 \\
& Lopera & 0.62 & 0.09 & 0.58 & 0.08 \\
& Los Baños & 0.16 & 0.08 & 0.49 & 0.13 \\
& Cortes & 0.30 & 0.07 & 0.67 & 0.06 \\
\midrule
\multirow{2}{6em}{CUEVAS CAMPO} & La Colonia & 0.32 & 0.04 & 0.51 & 0.05 \\
& Cuevas del Campo & 0.46 & 0.01 & 0.73 &  0.03 \\
\midrule
\multirow{5}{6em}{CULLAR} & Las Vertientes & 0.58 & 0.16 & 0.65 & 0.11 \\
& Charcón Higuera & 0.41 & 0.04 & 1.11 & 0.10 \\
& Charcón Nicolases & 0.63 & 0.27 & 0.67 & 0.19 \\
& Cúllar & 0.42 & 0.01 & 0.76 & 0.02 \\
& El Margen & 0.43 & 0.06 & 0.62 & 0.07 \\
\midrule
DARRO & Darro & 0.54 & 0.03 & 0.69 & 0.03 \\
\midrule
DEHESAS DE GUADIX & Dehesas de Guadix & 0.73 & 0.04 & 0.88 & 0.04 \\
\midrule
DIEZMA & Diezma & 0.74 & 0.04 & 1.16 & 0.05 \\
\midrule
FONELAS & Fonelas & 0.39 & 0.02 & 0.33 & 0.02 \\
\midrule
FREILA & Freila & 0.59 & 0.03 & 0.89 & 0.05 \\
\midrule
GALERA & Galera & 0.48 & 0.03 & 1.14 & 0.07 \\
\midrule
GOBERNADOR & Gobernador & 0.07 & 0.01 & 0.36 & 0.03 \\
\midrule
GOR & Gor & 0.60 & 0.06 & 2.34 & 0.24 \\
\midrule
GORAFE & Gorafe & 0.79 & 0.06 & 0.67 & 0.05 \\
\midrule
\multirow{6}{6em}{GUADIX} & Bácor-Olivar & 0.18 & 0.02 & 0.94 & 0.07 \\
& Estación de Guadix & 0.42 & 0.03 & 0.31 & 0.02 \\
& Belerda & 0.75 & 0.11 & 0.32 & 0.05 \\
& Guadix & 0.22 & 0.01 & 0.36 & 0.01 \\
& Hernán-Valle & 0.43 & 0.04 & 0.31 & 0.03 \\
& Polígono Industrial & 0.40 & 0.02 & 0.52 & 0.02 \\
\midrule
HUELAGO & Huélago & 0.47 & 0.04 & 0.35 & 0.03 \\
\midrule
\multirow{3}{6em}{HUESCAR} & La Parra & 0.27 & 0.04 & 0.67 & 0.09 \\
& Huéscar & 0.28 & 0.01 & 0.42 & 0.01 \\
& Barrio Nuevo & 0.30 & 0.01 & 0.39 & 0.02 \\
\midrule
JEREZ DEL MARQUESADO & Jerez del Marquesado & 0.54 & 0.05 & 0.66 & 0.05 \\
\midrule
LA CALAHORRA & La Calahorra & 0.53 & 0.04 & 0.50 & 0.04 \\
\midrule
\multirow{2}{6em}{LA PEZA} & Los Villares & 0.55 & 0.07 & 0.93 & 0.10 \\
& La Peza & 0.83 & 0.08 & 0.70 & 0.05 \\
\midrule
LANTEIRA & Lanteira & 0.39 & 0.03 & 0.60 & 0.05 \\
\midrule
LUGROS & Lugros & 0.41 & 0.05 & 0.99 & 0.12 \\
\midrule
MARCHAL & Marchal & 0.33 & 0.03 & 0.49 & 0.03 \\
\midrule
\multirow{2}{6em}{MORELABOR} & Moreda & 0.15 & 0.01 & 0.33 & 0.03 \\
& Laborcillas & 0.46 & 0.04 & 0.39 & 0.03 \\
\midrule
ORCE & Orce & 0.41 & 0.02 & 0.39 & 0.02 \\
\midrule
PEDRO MARTÍNEZ & Pedro Martínez & 0.41 & 0.02 & 0.70 & 0.04 \\
\midrule
POLICAR & Polícar & 0.39 & 0.05 & 0.70 & 0.10 \\
\midrule
\multirow{3}{6em}{PUEBLA DE DON FADRIQUE} & Almaciles & 0.29 & 0.02 & 0.27 & 0.02 \\
& P. de D. Fadrique & 0.31 & 0.01 & 0.38 & 0.02 \\
& & & & & \\
\midrule
PURULLENA & Purullena & 0.59 & 0.03 & 0.88 & 0.05 \\
\midrule
\multirow{2}{6em}{VALLE DEL ZALABI} & Alcudia de Guadix & 0.36 & 0.03 & 0.57 & 0.04 \\
& Esfiliana & 0.54 & 0.10 & 0.35 & 0.05 \\
\midrule
VILLANUEVA DE LAS TORRES & Vva. de las Torres & 0.37 & 0.02 & 0.43 & 0.03 \\
\midrule
ZUJAR & Zújar & 0.52 & 0.02 & 0.67 & 0.02 \\
\midrule
\multirow{5}{6em}{PRIVATE INSTALLATIONS} & Quarry (Darro) & 0.57 & 0.06 & 0.62 & 0.08 \\
& Andasol & 0.27 & 0.01 & 0.32 & 0.01 \\
& Logistics Centre & 0.47 & 0.02 & 0.73 & 0.04 \\
& Service area & 0.43 & 0.04 & 0.76 & 0.07 \\
& CELSUR & 0.44 & 0.02 & 0.67 & 0.04 \\

\bottomrule
\end{longtable}
\end{ThreePartTable}

\clearpage

\section{Results of cluster analysis (k-means)}\label{A:3}

\begin{ThreePartTable}

\begin{TableNotes}
\scriptsize
\item Group assignment in k-means partitions. $K_1$: group according to type of lamp inventoried in street lighting; $K_2$: group according to B/G colour index and VIIRS differences between 2021 and 2012; $K_3$: group according to colour indexes in 2021; $K_4$: group according to colour indexes difference between 2021 and 2012. The numbers in brackets next to the names of the population centres are the same as those shown in Figure \ref{fig:05}.
\end{TableNotes}

\scriptsize
\begin{longtable}{@{}l|ll|l|cc@{}}
\caption{K-means cluster analysis results: group assignment in k-means partitions} 
\label{tab:kmeans} \\
\toprule
\textbf{MUNICIPALITY} & \textbf{$K_1$} & \textbf{$K_2$} & \textbf{CENTRE} & \textbf{$K_3$} & \textbf{$K_4$} \\
\midrule
\endfirsthead

\multicolumn{5}{c}{{\bfseries \tablename\ \thetable{}:}  continuation}\\
\addlinespace[0.9ex]
\toprule
\textbf{MUNICIPALITY} & \textbf{$K_1$} & \textbf{$K_2$} & \textbf{CENTRE} & \textbf{$K_3$} & \textbf{$K_4$} \\
\midrule    
\endhead

\midrule
\endfoot
\bottomrule
\insertTableNotes
\endlastfoot

ALAMEDILLA & 2 & 1 & (62) Alamedilla & 3 & 3 \\
\midrule
ALBUÑÁN & 3 & 1 & (51) Albuñán & 3 & 3 \\
\midrule
ALDEIRE & 1 & 1 & (38) Aldeire & 2 & 2 \\
\midrule
ALICÚN DE ORTEGA & 3 & 1 & (67) Alicún de Ortega & 3 & 3 \\
\midrule
ALQUIFE & 1 & 1 & (52) Alquife & 3 & 3 \\
\midrule
\multirow{4}{6em}{BAZA} & \multirow{4}{2em}{-} & \multirow{4}{2em}{-} & (13) Río de Baza & 2 & 3 \\
& & & (14) Baza & 3 & 3 \\
& & & (20) Baúl & 3 & 3 \\
& & & (79) P.I. el Baico & 2 & 2 \\
\midrule
BEAS DE GUADIX & 3 & 1 & (42) Beas de Guadix & 3 & 2 \\
\midrule
BENALÚA & 3 & 3 & (59) Benalúa & 3 & 3 \\
\midrule
\multirow{2}{6em}{BENAMAUREL} & \multirow{2}{2em}{1} & \multirow{2}{2em}{1} & (17) Benamaurel & 2 & 3 \\
& & & (18) San Marcos & 2 & 3 \\
\midrule
\multirow{3}{6em}{CANILES} & \multirow{3}{2em}{2} & \multirow{3}{2em}{2} & (15) Caniles & 3 & 3 \\
& & & (19) La Vega & 2 & 3 \\
& & & (25) Los Gallardos & 3 & 3 \\
\midrule
CASTILLÉJAR & 1 & 1 & (30) Castilléjar & 3 & 3 \\
\midrule
\multirow{3}{6em}{CASTRIL} & \multirow{3}{2em}{1} & \multirow{3}{2em}{1} & (9) Castril & 3 & 2 \\
& & & (10) Almontaras & 2 & 1 \\
& & & (11) Fátima & 3 & 3 \\
\midrule
COGOLLOS DE GUADIX & 3 & 1 & (50) Cogollos de Guadix & 3 & 3 \\
\midrule
\multirow{4}{6em}{CORTES DE BAZA} & \multirow{4}{2em}{3} & \multirow{4}{2em}{4} & (6) La Teja & 1 & 1 \\
& & & (8) Los Laneros & 2 & 2 \\
& & & (16) Cortes de Baza & 2 & 2 \\
& & & (21) Campo Cámara & 3 & 3 \\
\midrule
\multirow{4}{6em}{CORTES Y GRAENA} & \multirow{4}{2em}{-} & \multirow{4}{2em}{-} & (44) Graena & 3 & 3 \\
& & & (45) Lopera & 2 & 2 \\
& & & (46) Los Baños & 3 & 3 \\
& & & (47) Cortes & 3 & 3 \\
\midrule
\multirow{3}{6em}{CUEVAS DEL CAMPO} & \multirow{3}{2em}{1} & \multirow{3}{2em}{3} & (7) La Colonia & 3 & 3 \\
& & & (12) Cuevas del Campo & 2 & 2 \\
& & & & & \\
\midrule
\multirow{5}{6em}{CÚLLAR} & \multirow{5}{2em}{1} & \multirow{5}{2em}{3} & (1) Las Vertientes & 2 & 2 \\
& & & (26) Charcón Higuera & 1 & 1 \\
& & & (27) Charcón Nicolases & 2 & 2 \\
& & & (28) Cúllar & 3 & 3 \\
& & & (29) El Margen & 2 & 3 \\
\midrule
DARRO & 2 & 4 & (71) Darro & 2 & 2 \\
\midrule
DEHESAS DE GUADIX & 2 & 4 & (64) Dehesas de Guadix & 2 & 2 \\
\midrule
DIEZMA & 2 & 4 & (35) Diezma & 2 & 2 \\
\midrule
FONELAS & 3 & 1 & (60) Fonelas & 3 & 3 \\
\midrule
FREILA & 2 & 4 & (4) Freila & 2 & 1 \\
\midrule
GALERA & 1 & 1 & (31) Galera & 1 & 1 \\
\midrule
GOBERNADOR & 3 & 1 & (69) Gobernador & 3 & 3 \\
\midrule
GOR & - & - & (2) Gor & 1 & 1 \\
\midrule
GORAFE & 3 & 1 & (65) Gorafe & 2 & 3 \\
\midrule
\multirow{6}{6em}{GUADIX} & \multirow{6}{2em}{-} & \multirow{6}{2em}{-} & (3) Bácor-Olivar & 3 & 3 \\
& & & (53) Estación de Guadix & 3 & 2 \\
& & & (57) Belerda & 2 & 2 \\
& & & (66) Guadix & 3 & 3 \\
& & & (72) Hernán-Valle & 3 & 3 \\
& & & (76) Polígono Industrial & 3 & 3 \\
\midrule
HUÉLAGO & 3 & 1 & (58) Huélago & 3 & 3 \\
\midrule
\multirow{3}{6em}{HUÉSCAR} & \multirow{3}{2em}{3} & \multirow{3}{2em}{2} & (23) La Parra & 3 & 2 \\
& & & (32) Huéscar & 3 & 3 \\
& & & (34) Barrio Nuevo & 3 & 3 \\
\midrule
JEREZ DEL MARQUESADO & 1 & 1 & (49) Jerez del Marquesado & 2 & 2 \\
\midrule
LA CALAHORRA & 3 & 1 & (56) La Calahorra & 3 & 3 \\
\midrule
\multirow{2}{6em}{LA PEZA} & \multirow{2}{2em}{1} & \multirow{2}{2em}{1} & (36) Los Villares &  2 & 2 \\
& & & (41) La Peza & 2 & 2 \\
\midrule
LANTEIRA & 3 & 1 & (37) Lanteira & 3 & 3 \\
\midrule
LUGROS & 1 & 1 & (39) Lugros & 1 & 1 \\
\midrule
MARCHAL & 3 & 1 & (43) Marchal & 3 & 3 \\
\midrule
\multirow{2}{6em}{MORELÁBOR} & \multirow{2}{2em}{3} & \multirow{2}{2em}{1} & (68) Moreda & 3 & 3 \\
& & & (70) Laborcillas & 2 & 2 \\
\midrule
ORCE & 3 & 4 & (33) Orce & 3 & 2 \\
\midrule
PEDRO MARTÍNEZ & 2 & 4 & (61) Pedro Martínez & 3 & 3 \\
\midrule
POLÍCAR & 2 & 1 & (40) Polícar & 3 & 3 \\
\midrule
\multirow{3}{6em}{PUEBLA DE DON FADRIQUE} & \multirow{3}{2em}{3} & \multirow{3}{2em}{2} & (22) Almaciles & 3 & 3 \\
& & & (24) P. de D. Fadrique & 3 & 3 \\
& & & & & \\
\midrule
PURULLENA & 2 & 3 & (48) Purullena & 2 & 2 \\
\midrule
\multirow{2}{6em}{VALLE DEL ZALABÍ} & \multirow{2}{2em}{3} & \multirow{2}{2em}{1} & (54) Alcudia de Guadix & 3 & 3 \\
& & & (55) Esfiliana & 2 & 2 \\
\midrule
VILLANUEVA DE LAS TORRES & 3 & 1 & (63) Vva. de las Torres & 3 & 3 \\
\midrule
ZÚJAR & 3 & 4 & (5) Zújar & 2 & 2 \\
\midrule
\multirow{5}{6em}{PRIVATE INSTALLATIONS} & \multirow{5}{2em}{-} & \multirow{5}{2em}{-} & (73) Quarry (Darro) & 2 & 2 \\
& & & (74) Andasol & 3 & 3 \\
& & & (75) Logistics Centre & 2 & 3 \\
& & & (77) Service area & 2 & 3 \\
& & & (78) CELSUR & 2 & 3 \\

\bottomrule
\end{longtable}
\end{ThreePartTable}
 
\end{document}